\pdfoutput=1 

\documentclass[a4paper,aps,prl,onecolumn,notitlepage,superscriptaddress,showpacs,amsmath,amssymb]{revtex4-1}

\usepackage{geometry}
\usepackage{graphicx}
\usepackage{dcolumn}
\usepackage{bm}
\usepackage[usenames,dvipsnames]{xcolor} 
\usepackage[colorlinks,citecolor=MidnightBlue,linkcolor=Orange,urlcolor=MidnightBlue]{hyperref}
\usepackage[mathlines]{lineno}
\usepackage{color}
\usepackage{MnSymbol}
\usepackage{times} 

\begin{document}
	
\title{Reconfigurable Artificial Microswimmers with Internal Feedback}

\author{L. Alvarez}
\affiliation{Laboratory for Soft Materials and Interfaces, Department of Materials, ETH Zurich, 8093 Zurich, Switzerland}%
\author{M. A. Fernandez-Rodriguez}%
\affiliation{Laboratory for Soft Materials and Interfaces, Department of Materials, ETH Zurich, 8093 Zurich, Switzerland}%
\affiliation{Biocolloid and Fluid Physics Group, Applied Physics Department, Faculty of Sciences, University of Granada, 18071 Granada, Spain}%
\author{A. Alegria}%
\author{S. Arrese-Igor}%
\affiliation{Centro de F\'isica de Materiales (SCIC-UPV/EHU), Materials Physics Center, 20018 San S\'ebastian, Spain}
\author{K. Zhao}
\affiliation{Laboratory for Soft Materials and Interfaces, Department of Materials, ETH Zurich, 8093 Zurich, Switzerland}%
\author{M. Kr\"oger}
\affiliation{Polymer Physics, Department of Materials, ETH Zurich, 8093 Zurich, Switzerland}%
\author{Lucio Isa}
\affiliation{Laboratory for Soft Materials and Interfaces, Department of Materials, ETH Zurich, 8093 Zurich, Switzerland}%

	\begin{abstract}
	Micron-size self-propelling particles are often proposed as synthetic models for biological microswimmers, yet they lack internally regulated adaptation, which is central to the autonomy of their biological counterparts. Conversely, adaptation and autonomy can be encoded in larger-scale soft-robotic devices, but transferring these capabilities to the colloidal scale remains elusive. Here, we create a new class of responsive microswimmers, powered by induced-charge electrophoresis, which can adapt their motility to external stimuli via an internal feedback. Using sequential capillary assembly, we fabricate deterministic colloidal clusters comprising soft thermoresponsive microparticles, which, upon spontaneous reconfiguration, induce motility changes, such as adaptation of the clusters' propulsion velocity and reversal of its direction. We rationalize the response in terms of a coupling between self-propulsion and variations of particle shape and dielectric properties. Harnessing those allows for strategies to achieve local dynamical control with simple illumination patterns, revealing exciting opportunities for the development of new tactic active materials.
	\end{abstract}
	\maketitle
	
	The ubiquity and success of motile bacteria is strongly coupled to their ability to autonomously adapt to different environments as they can reconfigure their shape, metabolism and motility via internal feedback mechanisms \cite{Hamadeh2011,Baker2006,berg2008coli,Katuri2017,Frangipane2018, Arlt2018, Mathijssen2019}. Realizing artificial microswimmers with similar adaptation capabilities and autonomous behavior will have disruptive effects on technologies from optimal transport to sensing and microrobotics \cite{Bechinger2016}. Existing approaches to impart adaptation at the colloidal scale mostly rely on external feedback, either to regulate motility via the spatio-temporal modulation of the propulsion velocity and direction \cite{Lozano2016, lavergne2019group, Khadka2018, Sprenger2019, FernandezRodriguez2020} or to induce shape changes via the same magnetic or electric fields \cite{Han2017,Shields2017,Yang18186}, which are also driving the particles. However, the necessity of external intervention conceptually contradicts the pursuit of autonomous behavior \cite{Ebbens2016}.  
	
	It is therefore essential to endow artificial microswimmers with an internal feedback mechanism that regulates motility in response to stimuli, which are decoupled from the source of propulsion. A promising route to achieve this goal is to exploit the coupling between particle shape and motility. Efficient switching between different propulsion states can for instance be reached by the spontaneous aggregation of symmetry-breaking active clusters of varying geometry \cite{Soto2014, Niu2018, Ma2015a, Wang2020}, albeit this process does not have the desired deterministic control. On the contrary, designing colloidal clusters with fixed shapes and compositions offers fine control on motility \cite{Ni2017,Wang2019}, but lacks adaptation. Although progress on reconfigurable robots at the sub-millimeter scale has been made \cite{Wenqi2018,Huang2019}, downscaling this concept to the colloidal level demands an innovative fabrication design. Shape-shifting colloidal clusters reconfiguring along a predefined pathway in response to local stimuli \cite{Dou2019} would combine both characteristics, holding promise to approach the vision of realizing fully adaptive and autonomous artificial microswimmers. 
	
	To achieve this, we fabricate geometrically and compositionally asymmetric colloidal clusters containing both polystyrene (PS) microparticles ($2$ \textmu{}m $\diameter$, fluorescent Ex/Em - 530/607 nm) and soft thermo-responsive microgels (poly-isopropylacrylamide-co-acrylic acid, PNIPAM-co-AAC, hydrodynamic diameter $1.7$ \textmu{}m) using sequential Capillarity-Assisted Particle Assembly (sCAPA) \cite{Ni2016, Ni2017}.
	
\begin{figure*}[t]\includegraphics[width=0.7\textwidth]{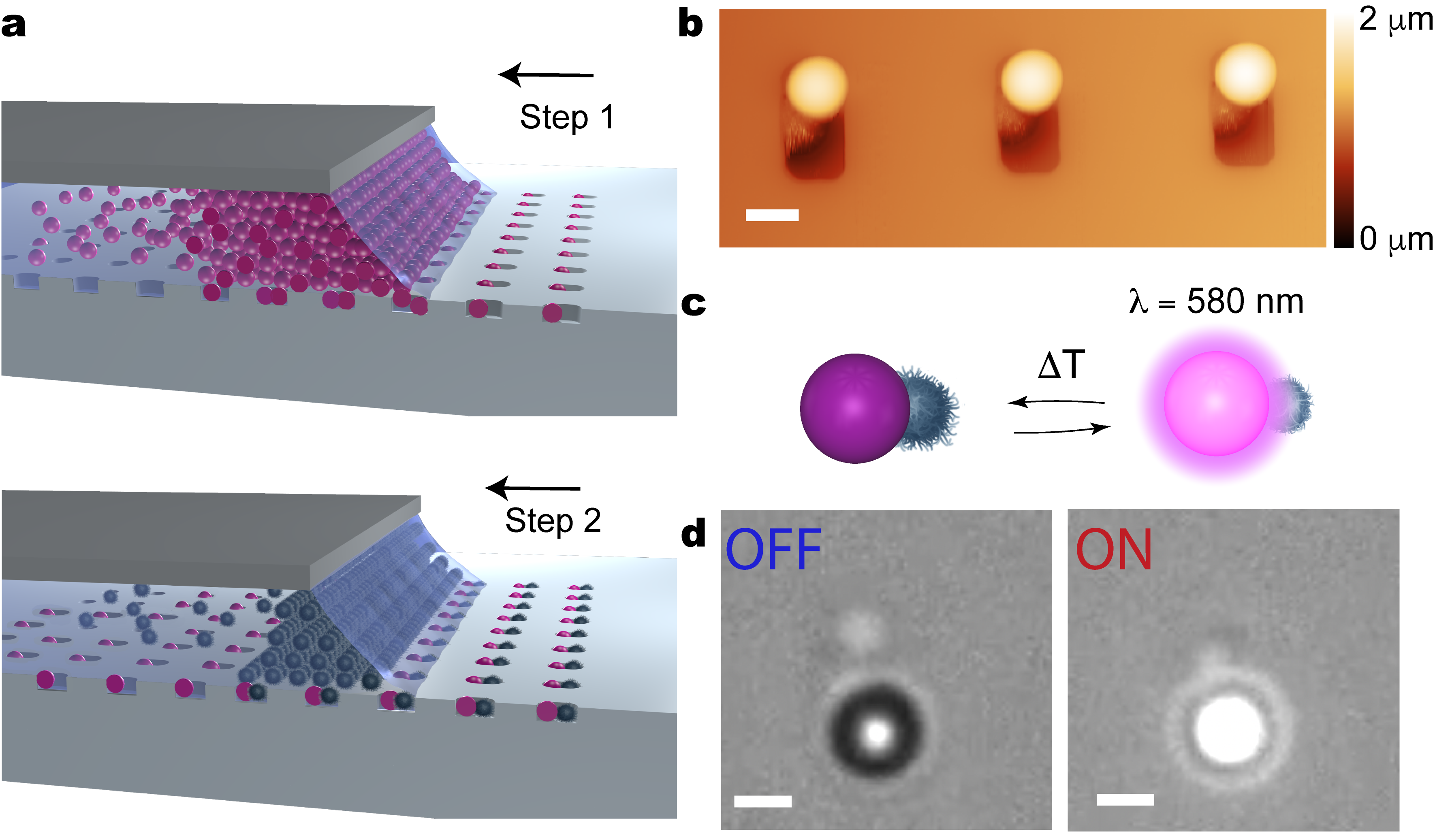}
		\caption{\textbf{Fabrication of thermo-responsive reconfigurable dumbbells}.~\textbf{a}, Scheme of the deposition of polystyrene particles (step 1) and PNIPAM-co-AAC microgels (step 2) onto a PDMS template patterned with $2\times 4$ \textmu{}m$^2$ size traps. The arrow indicates the direction of the deposition. Particles are accumulated at the meniscus of the moving droplet and deposited into the traps via capillary forces. \textbf{b}, AFM image in water showing PS-microgel dumbbells in the traps before harvesting. The bright spheres are the PS colloids and the darker ones are the swollen microgels. \textbf{c-d}, Schematic representation \textbf{c} and optical micrographs \textbf{d} of the light-driven reconfiguration of a dumbbell. Illumination close to the absorption wavelength of the PS particle ($\lambda=580$ nm) with a power density $\rho_{\rm FL}=0.4$ mW/mm$^2$, locally increases the temperature above the VPTT of the microgel, causing it to deswell and correspondingly induce a variation of the dumbbell's geometry and dielectric properties. The transition is reversible upon removing the incident light. Scale bars: 2 \textmu{}m. }
		\label{fig:figure1}
	\end{figure*}

	Briefly, we assemble both particles onto a substrate that has traps with a geometry corresponding to the target shape of the final cluster. We first deposit the PS particles and, in a second step, we add the microgels, which fill the free space in the traps left after the first deposition (Fig.~\ref{fig:figure1}a). The trap-filling process is enabled by the capillary force exerted by the meniscus of an evaporating droplet of the colloidal suspensions driven over the substrate at a controlled speed.  The formed PS-microgel clusters are then thermally sintered to ensure their mechanical integrity upon harvest and transferred to the experimental cell (more details in Materials and Methods). We achieve the controlled formation of cluster populations with different geometries on different regions of the same template, e.g. dumbbells and L-shapes, defined by the local orientation of the moving meniscus relative to the trap orientation. The highest population is composed by dumbbells (see Fig.~\ref{fig:figure1}b and Supplementary Fig.~S1). 
	
	The incorporation of the PNIPAM-co-AAC microgels endows the dumbbells with a reversible temperature response. These microgels exhibit a volume phase transition (VPT) in water at a temperature $T =$ 32\textdegree{}C. Upon crossing the VPTT, the microgels reduce their volume by approximately 80\%, almost double the magnitude of their zeta potential ($\zeta$) (Supplementary Fig.~S2), and drastically change their dielectric properties. In particular, they show a 50\% drop of their dielectric permittivity $\epsilon'$ (Supplementary Fig.~S2). Over the same $T$ range, the corresponding properties of the PS particles remain practically unchanged. However, these particles have the function to act as the vehicle to induce the reconfiguration of the microgels. Illuminating the dye-loaded PS particles using light with a wavelength close to their excitation maximum ($\lambda = 580$ nm), causes local heating due to non-radiative energy dissipation. By controlling the fluorescence power density ($\rho_{\rm FL}$), temperature can locally increase above 32\textdegree{}C (Supplementary Fig.~S3), triggering the reconfiguration of the microgels attached onto the PS particles (Fig.~\ref{fig:figure1}c,d and Supplementary Movie S1). Therefore, the integrated PS-microgel cluster contains the essential elements of an autonomous adaptive two-state system: the PS particles convert an external light intensity signal into heat causing the transition of the microgels from a swollen to a collapsed state, which can be reversed by reducing the illumination. If such reconfiguration is coupled to self-propulsion, the result is a reconfigurable microswimmer with internal feedback, sensing and adaptive capabilities. 
	
\begin{figure*}[t]
		\includegraphics[width=0.8\textwidth]{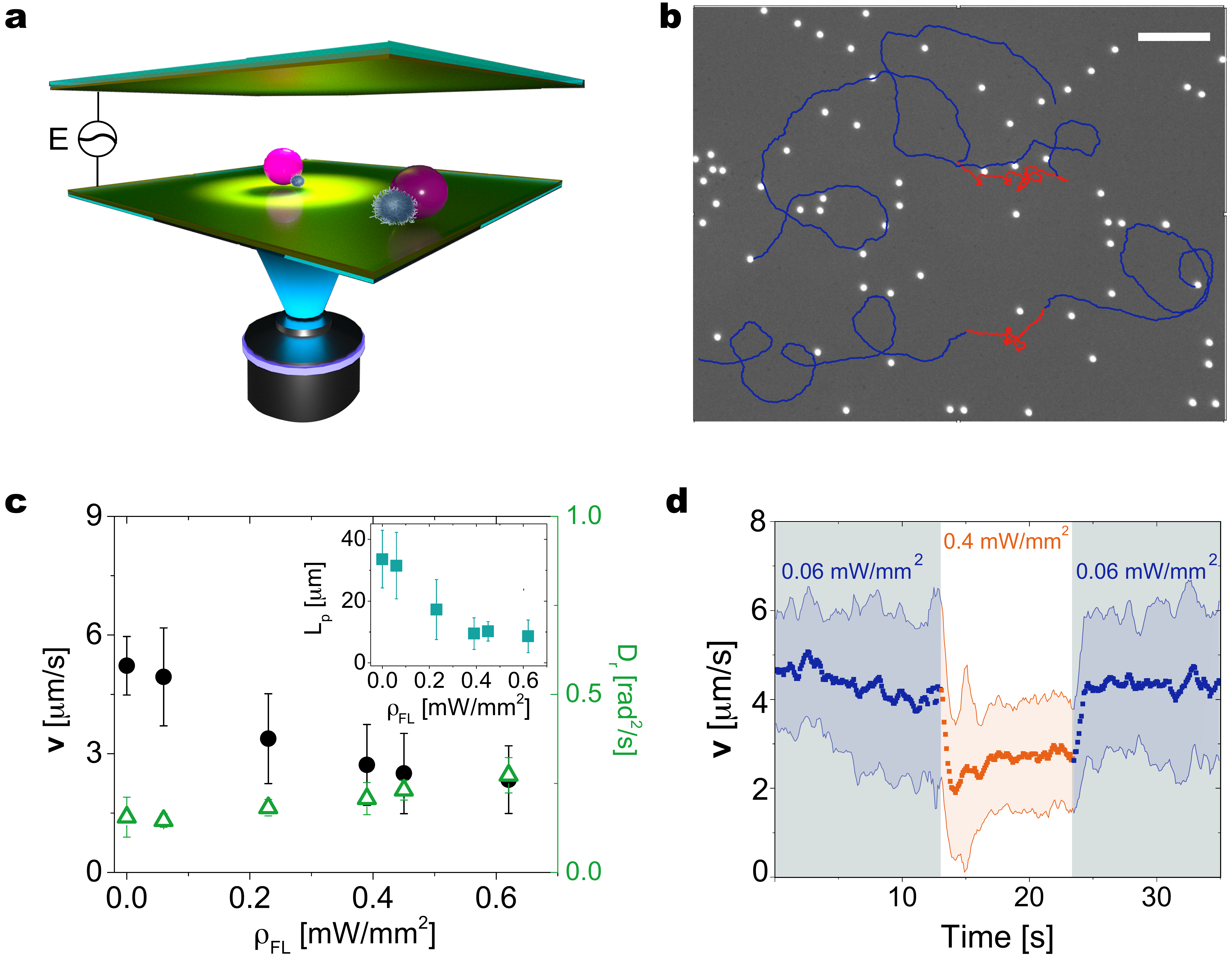}
		\caption{\textbf{Experimental realization of adaptive active dumbbells}.~\textbf{a}, Schematic representation of the experimental cell, illustrating the transverse AC electric field E and the local illumination with fluorescent light, respectively generating motion and causing particle reconfiguration. \textbf{b}, Examples of particle trajectories switching $\rho_{\rm FL}$ between 0.06 (blue trajectory) and 0.4 $\textrm{mW/mm}^2$ (red trajectory). At $\rho_{\rm FL} = 0.4$ mW/mm$^2$, the PS particle heats the microgel above its VPTT, causing a motility change. \textbf{c}, $v$ (solid circles) and $D_{\rm r}$ (open triangles) as a function of illumination power density $\rho_{\rm FL}$. The inset represents the persistence length ($L_p$) of the particles' trajectories as a function of $\rho_{\rm FL}$. \textbf{d}, Particle velocity as a function of time for two levels of $\rho_{\rm FL}= 0.06$ (shaded areas)  and 0.4 (white area)  mW/mm$^2$. Scale bar represents 20 \textmu{}m. Error bars in all cases indicate the standard deviation of the data.}
		\label{fig:figure2}
	\end{figure*}

	
	In this work, such coupling is reached by driving self-propulsion via induced-charge electro-phoresis (ICEP) \cite{PhysRevLett.92.066101}. In ICEP, colloidal particles self-propel over an electrode, which generates a spatially uniform transverse AC electric field, due to locally unbalanced electrohydrodynamic flows created by an asymmetry of the particle's shape and dielectric properties. In our experiments, we confine an aqueous suspension of the reconfigurable clusters between two conductive transparent surfaces separated by a spacer of thickness $h$. Upon applying an AC voltage with peak-to-peak amplitude $V_{\rm pp}$ = 4--5 V  at a frequency $f= 1$ kHz, the clusters self-propel with a swimming velocity $v \propto (V_{\rm pp}/h)^2$, leading to velocities that vary between $3$--$6$ \textmu{}m/s at room temperature. We track the motion of the particles by means of an inverted microscope (40$\times$ magnification objective at 10 fps) mixing transmission and epifluorescence illumination (Fig.~\ref{fig:figure2}a), where the modulation of the latter source encodes the input signal for particle reconfiguration. We first focus on the dynamics of dumbbells and later move to L-shaped particles.  
	
	The example trajectories depicted in Fig.~\ref{fig:figure2}b show that the motility of self-propelling dumbbells adapts to the level of fluorescence illumination. By independently measuring $v$ and the rotational diffusivity $D_r$ as a function of $\rho_{\rm FL}$ (Fig.~\ref{fig:figure2}c, details in the Methods Section), we observe that $D_r$ increases due to (i) particle size changes following the microgel collapse, and (ii) heat-induced local viscosity changes (Supplementary Fig.~S5). Correspondingly, the velocity exhibits a counter-intuitive marked reduction up to $\rho_{\rm FL}=0.4$ mW/mm$^2$, after which it remains practically constant. In particular, we see that, upon switching from a low ($\rho_{\rm FL}=0.06$ mW/mm$^2$, blue trajectory) to a high ($\rho_{\rm FL}=0.4$ mW/mm$^2$, red trajectory) fluorescence power density, the persistence length of the trajectories, defined as $L_p= v/D_r$, experiences a three-fold decrease, for an input illumination that induces the microgel collapse (inset to Fig.~\ref{fig:figure1}c). This behavior resonates with light-responsive bacteria, which adapt the way in which they explore their surroundings based on light signals \cite{Arlt2018, Frangipane2018}. A systematic quantification carried out over ~400 particles shows that the velocity changes are fully reversible (Supplementary Movies S2 and S3) and that the adaptation to illumination, and hence temperature, variations has a characteristic response time of a few seconds  (Fig.~\ref{fig:figure1}d). This finite response time may be related to the timescale of the reconfiguration of microgels adsorbed on the PS particle surface (Supplementary Movie S1). 
	
\begin{figure*}[h!]
		\includegraphics[width=1\textwidth]{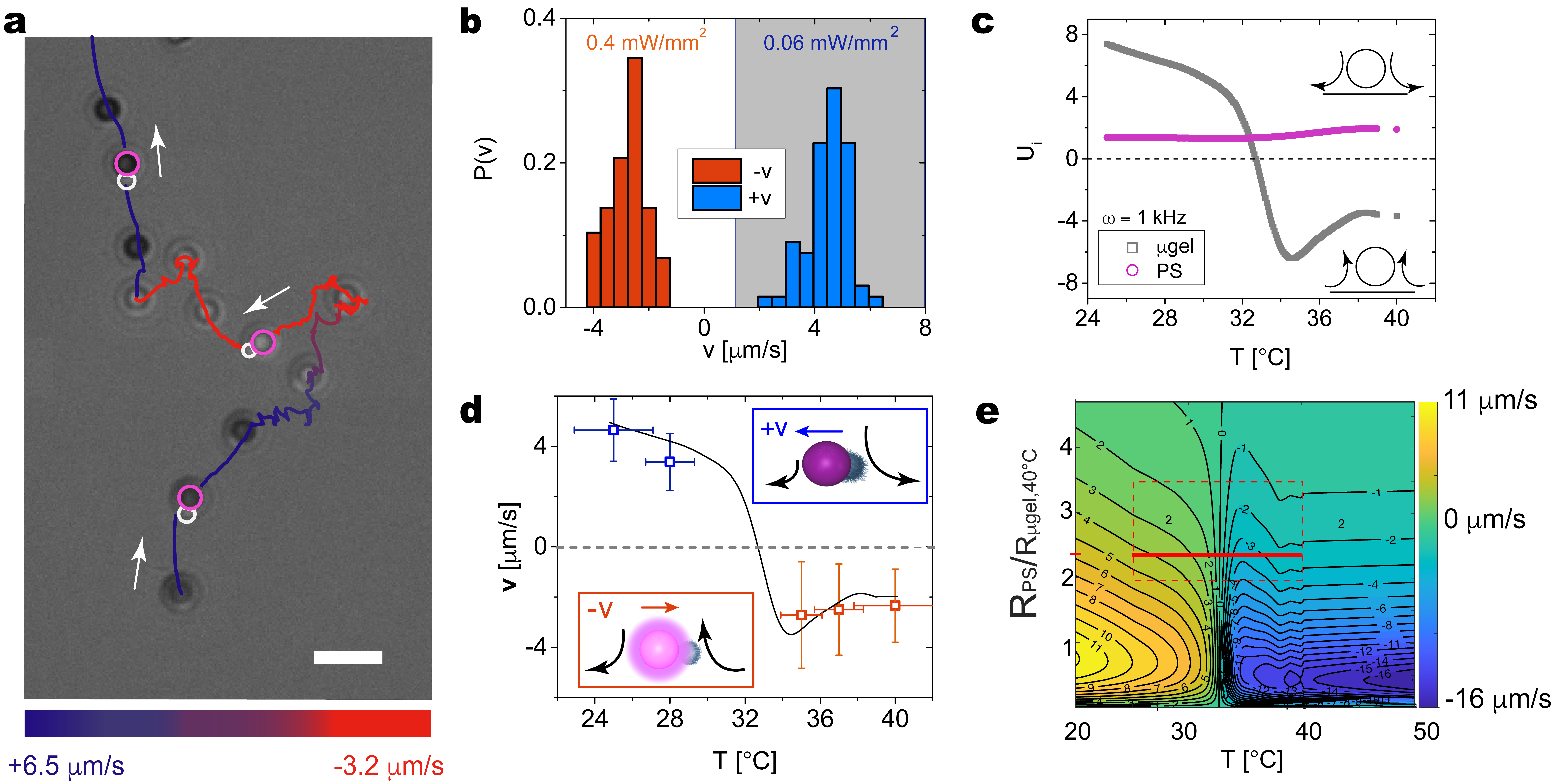}
		\caption{\textbf{Feedback between propulsion mechanism and particle reconfiguration}.~ \textbf{a}, Particle trajectory (63$\times$ magnification) showing that at $\rho_{\rm FL}=0.06$ mW/mm$^2$ (blue) the particle swims toward the PS lobe ($+v$), and at $\rho_{\rm FL}=0.4$ mW/mm$^2$ (red) it  changes direction and swims with the microgel in front ($-v$). The pink and white circle indicate the position of the PS particle and microgel, respectively. \textbf{b}, Histogram of particle velocity (45 particles) at low $\rho_{\rm FL}$ (0.06 mW/mm$^2$ - grey area) and high $\rho_{\rm FL}$ (0.4 mW/mm$^2$ - white area). \textbf{c}, EHD flow velocities $U_i$ for each particles calculated from Eq.\ \eqref{eq:EHD}. The schematic indicates the direction of the EHD flows, either positive (repulsive) or negative (attractive).~ \textbf{d}, Experimental values (symbols) and theoretical prediction (solid line) of dumbbell velocity as a function of temperature. The inset schemes indicate the EHD flows and final propulsion direction. y-error bars are the standard deviation of the velocity calculated for 45 dumbbells at each temperature. x-error bars correspond to the uncertainty in experimentally determining $T$. \textbf{e}, Theoretical prediction of dumbbell velocity $v$ as a function of $T$ and size ratio between the PS sphere and the collapsed microgel (at 40\textdegree{}C). The plot is obtained from Eq.\ \eqref{eq:EHD} using the experimental temperature-dependent system properties (particle size and dielectric properties, solvent viscosity) as inputs and a single-temperature independent prefactor. The dashed red box highlights the experimentally accessible range of $T$ and size ratios (the range of size ratios is estimated from the error bars in Fig.\ \ref{fig:figure3}d). The red solid horizontal line  corresponds to the black solid curve in Fig.\ \ref{fig:figure3}d.}
		\label{fig:figure3}
	\end{figure*}

	A closer inspection of the time-dependence of the motility changes reveals a rich dynamical response, hinged on the feedback between the propulsion mechanism and the reconfiguration of the particle properties. At higher magnifications (63$\times$ objective at 10 fps), both the PS and the microgel lobe can be distinguished during self-propulsion of the dumbbells (Fig.~\ref{fig:figure3}a). At low $\rho_{\rm FL}$, we observe that all particles swim with the PS lobe in front ($+v$). At high $\rho_{\rm FL}$ (0.4 mW/mm$^2$), we instead observe that the particles invert their direction of motion and start swimming at a different speed with the microgel lobe in front ($-v$) (Fig.~\ref{fig:figure3}a-b, Supplementary Movie S4).
	
	
	This dynamical response derives from the changes of the local EHD flows around a PS-microgel dumbbell following the microgel collapse. The fluid velocity as a function of distance $r$ away from the surface of either spherical particle is given by  
	\begin{equation}
	\label{eq:EHD} 
	U_i = \frac{C K_i''}{\eta}\frac{3(r/R_i)}{2 \left[ 1+ (r/R_i)^2\right]^{5/2}},
	\end{equation}
	where $\eta$ is the solvent viscosity, $R_i$ the particle size, $K_i''$ the imaginary part of the Clausius-Mossotti factor and $i$ indicates either the PS particle or the microgel (see SI for detailed explanation) \cite{CMFactor2017,FuduoMa2014}. These quantities are temperature-dependent (Supplementary Fig.\ S8) and $C$ is a constant prefactor used as single fitting parameter \cite{Ma2015a}. We used dielectric spectroscopy to measure the dielectric constant $\epsilon'$ and conductivity $\sigma'$ of a microgel dispersion at a frequency of 1 kHz for temperatures between 25--40\textdegree{}C, corresponding to the conditions of our experiments. From these data, we calculate $K_\textrm{\textmu{}gel}''$, while $K_\textrm{PS}''$ was obtained from literature values \cite{ERMOLINA2005419}.
	These inputs are used to predict $U_{i}$ \cite{Ma2015}, where we see that $U_\textrm{\textmu{}gel}$ decreases and inverts its direction as $T$ grows. Correspondingly, $U_\textrm{PS}$ exhibits only a slight increase, due a reduction of the fluid viscosity in the vicinity of the heated particle (Fig.\ \ref{fig:figure3}c).
	
	These predictions are further supported by direct visualization of the EHD flows around each lobe using fluorescent tracers ($750$~nm) while applying the actuating AC field and changing the light conditions (Supplementary Fig.\ S9a) \cite{Ma2015,Yang2019,Campbell2019}. At low light intensities, the overall behavior is dominated by the repulsive flows around the microgel surface, causing the dumbbell to propel with the PS lobe in front. When the microgel shrinks at higher light intensities, the EHD flows close to its surface changes sign and is strongly reduced.  This causes an inversion of the propulsion direction, where the now stronger repulsive flows generated by the PS lobe propel the dumbbell with the microgel in front. 
	
	Combining the measured $T$-dependent properties for each particle, the dumbbell velocity $v(T)$ can, in first approximation, be obtained as a linear combination of the two values of $U_i$ as $v(T) = [U_\textrm{\textmu{}gel}(T) R_\textrm{PS} + U_\textrm{PS}(T) R_\textrm{\textmu{}gel}(T)] / [R_\textrm{\textmu{}gel}(T)+R_\textrm{PS}]$. Given the behavior of each single lobe, the dumbbell velocity changes both sign and magnitude when crossing the VPTT, as shown in Fig.\ \ref{fig:figure3}d. This theoretical prediction can be extended to different size ratios between the dumbbell's lobes, as shown in Fig.\ \ref{fig:figure3}e, with the solid red line corresponding to the $T$-range for our experiments shown in Fig.\ \ref{fig:figure3}d. This parametric representation offers promising guidelines for the design of dumbbells with a tailored velocity modulation and shows how the response is very sensitive to small variations close to the VPTT. Among the different parameters, it emerges that the changes in the dielectric properties of the microgels across their VPTT play the dominant role in regulating the dumbbell's dynamical response.
	
	\begin{figure*}[h!]
		\includegraphics[width=0.8\textwidth]{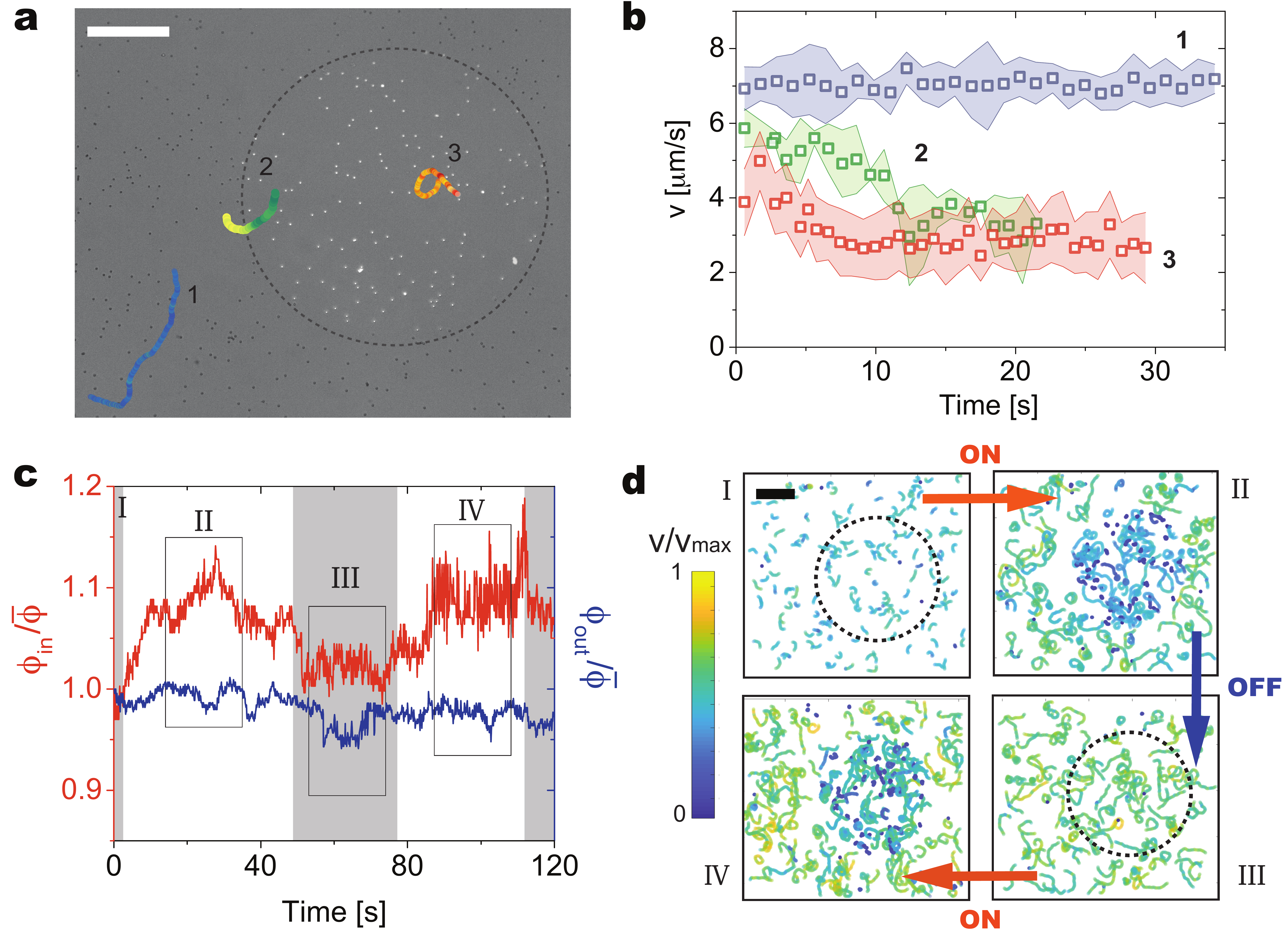}
		\caption{\textbf{Adapting swimming to light patterns}.~\textbf{a}, Combined transmission and epifluorescene micrograph of self-propelling dumbbells where the fluorescent light ($\rho_{\rm FL}=0.2$ mW/mm$^2$) is confined within the dashed circle by partly closing the microscope diaphragm. Characteristic trajectories of active particles remaining outside (light blue to dark blue) or inside (red to yellow) the illuminated region or crossing from one to the other (yellow to green). The trajectories are color-coded based on the instantaneous velocity. Upon entering the illuminated region, the dumbbell clearly slows down. Scale bar: 150 \textmu{}m. \textbf{b}, Particle velocities (absolute values) as a function of time for the three representative particles depicted in \textbf{a}, using the same colors. Time $t=0$ s corresponds to the fluorescent light being turned on. \textbf{c}, Particle number density inside $\phi_\textrm{in}$ (red - left axis) and outside the illuminated region $\phi_\textrm{out}$ (blue - right axis) versus time during on (white) and off (grey) cycles of the fluorescent light. The number density is normalized by the initial number density, $\bar{\phi}$, of a uniform particle distribution before turning on the fluorescent light. The data are produced by cumulating three independent experiments. \textbf{d}, Particle trajectories during the on-off cycles corresponding to the boxed regions in (c) color-coded by instantaneous velocity (averaged over 4 frames) over 50 frames (I) and 200 frames (II, III, IV). During the on cycles, the particles within the illuminated region significantly slow down and the persistence of their trajectories drops. During the off cycles they return to their original swimming behavior. The particles outside the illuminated region are not affected. Scale bars represent 40 \textmu{}m. }
		\label{fig:figure5}
	\end{figure*}
	
	The presence of light-driven reconfiguration enables us to modulate the particle swimming behavior using simple light patterns. By partly closing the diaphragm of the fluorescence illumination, we can create circular regions with controlled values of $\rho_{\rm FL}$ (Fig.~\ref{fig:figure5}a) and track the motion of active dumbbells inside or outside these regions, or crossing between the two. We find that particles remaining outside the illuminated region (blue data in Fig.~\ref{fig:figure5}b) consistently show a constant higher velocity than the ones inside the circular light pattern (red data), where $t =$ 0~s corresponds to the time when the fluorescence is turned on.  The particles inside the illuminated region show that the adaptation of the propulsion velocity has a characteristic time scale of a few seconds (see also Fig.~\ref{fig:figure1}d), analogously experienced by a particle that enters the illuminated region and progressively adapts to a different propulsion speed (green data).

	The position-dependent swimming imparted by the light patterns affects the collective behavior of ensembles of reconfigurable dumbbells. In particular, particle accumulation inside the illuminated regions can be elicited as a consequence of a progressive slowing down of particles entering the illuminated regions \cite{PhysRevLett.108.248101,PhysRevE.98.052606}. Cyclic on-and-off illumination causes a reversible increase of the particle number density inside the illuminated circle during the "on" times, which returns to the same value of the number density outside the circle when the fluorescent light is switched off (Fig.~\ref{fig:figure5}c). 
	The density changes for $\phi_\textrm{in}$ are clearly correlated to a decrease of particle propulsion speed and persistence of the trajectories within the illuminated region, as seen by the particle trajectories (Fig.~\ref{fig:figure5}d). The propulsion of the particles outside the illuminated circle remains unaffected. 
	
	\begin{figure*}[h!]
		\includegraphics[width=0.7\textwidth]{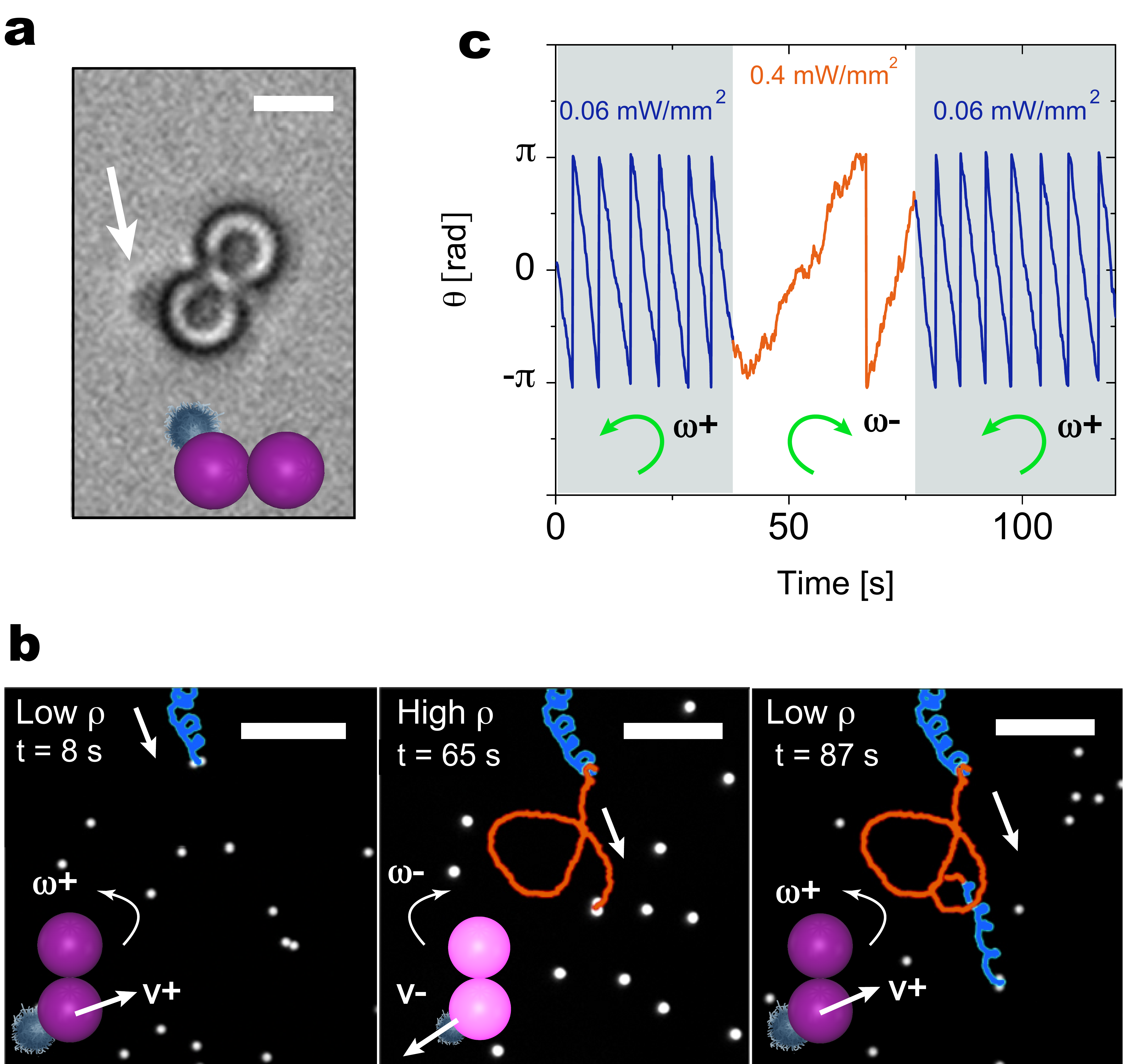}
		\caption{\textbf{Motion chirality control with L-shapes}.~\textbf{a}, Optical microscopy image of an L-shaped reconfigurable cluster comprising  two PS particles and one microgel in MilliQ water at room temperature. The arrow indicates the microgel position \textbf{b}, Trajectories of an L-shaped particle at two different levels of fluorescence light power density ($\rho_{\rm FL}=0.06$ mW/mm$^2$ - blue;  $\rho_{\rm FL}=0.4$ mW/mm$^2$ - red). The schematics in the insets show the propulsion direction relative to the PS-microgel combination as a function of $\rho_{\rm FL}$ and the corresponding angular velocity $\omega$. Upon changing illumination density, the helical trajectory changes chirality, due to propulsion reversal, and pitch, due to shape reconfiguration and velocity reduction. Scale bars: 5 \textmu{}m. \textbf{c}, Orientation angle as a function of time for a self-propelling L-shaped particle at different light illumination levels. Upon tuning $\rho_{\rm FL}$ up, the rotation changes direction and slows down. The change is fully reversible upon reducing the fluorescent light.}
		\label{fig:figure_L}
	\end{figure*}

	Finally, the coupling between internal reconfiguration and propulsion offers even more opportunities if the complexity of the cluster shape is enhanced. In nature, some bacteria have developed striking ways to correct their dynamics under flow by regulating helicoidal motion \cite{Matthinsen2019}.   Here, we show that sCAPA-fabricated L-shaped clusters comprising two PS spheres on the long arm and a PS-microgel dumbbell as the short arm (Fig.~\ref{fig:figure_L}a), show helical trajectories with a light-switchable chirality. L-shaped active particles are known to exhibit helical trajectories, whose pitch only depends on shape \cite{Kummel2013}. In our case, by changing the illumination conditions leading to the reconfiguration of the microgel, the L-shaped clusters simultaneously adapt their shape and propulsion velocity, both slowing down and inverting the propulsion direction relative to the orientation of the short arm of the L. As a result, the trajectory displayed in Fig.~\ref{fig:figure_L}b shows that both the pitch and the chirality of the motion changes from counter-clockwise $\omega^-$ to clockwise $\omega^+$, as confirmed by tracking the particle angular position over time (Fig.~\ref{fig:figure_L}c). 
	
	In conclusion, we present a new experimental realization of reconfigurable active colloids with internal feedback, coupling variations of shape and material properties. The modulation of persistence length, propulsion direction, and chirality emerges in direct response to external stimuli decoupled from the source of propulsion. This is possible through the incorporation of soft-responsive microgels in deterministic positions that endow the particles with internal degrees of freedom, which can be tailored to engineer both single-particle and collective response, such as tactic motion and localization. The ability to probe the environment and correspondingly adapt their dynamical behavior, brings us one step closer to mimicking the complex dynamical response of biological microswimmers and might set the basis for a next generation of autonomous active colloids.
	Future developments of this strategy are closely connected to progress in materials and microfabrication, in order to equip active particles with multiple responses and functionalities, downsizing the potential of soft robotics to the colloidal scale.
	\\\\\\\\\\\\\\
	
	\section*{Acknowledgements}
	The authors thank Peter Schurtenberger and Heiko Wolf for insightful discussions and Walter Richtering for providing the microgels. LI and LA acknowledge financial support from the Swiss National Science Foundation Grant PP00P2-172913/1 and the European Soft Matter Infrastructure (EUSMI) proposal number E190900328. 
	
	\section*{Author Contributions}
	Author contributions are defined based on the CRediT (Contributor Roles Taxonomy) and listed alphabetically. Conceptualization: L.A. and L.I.; Data curation: A.A., L.A., S.A.I. and K.Z.; Formal analysis: L.A., M.K. and K.Z.; Funding acquisition: L.I.; Investigation: L.A. and M.A.F-R.; Methodology: A.A., L.A., S.A.I., M.A.F-R., L.I., M.K. and K.Z.; Project administration: L.A. and L.I.; Software: L.A., M.K. and K.Z.; Supervision: L.I.; Validation: L.A. and M.A.F-R.;. Visualization: A.A., L.A., S.A.I. and  L.I.; Writing – original draft:L.A. and  L.I.; Writing – review and editing: A.A., L.A., S.A.I., M.A.F-R., L.I. and M.K..

	\section*{Additional Information}
	Supplementary Information is available for this paper. Correspondence and requests for materials should be addressed to laura.alvarez-frances@mat.ethz.ch and lucio.isa@mat.ethz.ch.

	\section*{Data availability statement}
	The data that support the findings of this study are available from the corresponding authors upon reasonable request.
	
	\section*{Code availability statement}
	The code used in this study is available from the corresponding authors upon reasonable request.

	\section{Materials and Methods}
	
	\subsection{Sequential capillarity-assisted particle assembly (sCAPA) and harvesting of reconfigurable colloidal clusters}
	
	Our active thermo-responsive clusters are formed by a combination of 2 \textmu{}m polystyrene spheres (PS, Microparticles Gmbh ) and PNIPAM-co-AAC microgel particles. The microgels have a hydrodynamic radius $R_\textrm{\textmu{}gel}= 0.85$\ \textmu{}m\ at 25\textdegree{}C, as measured in MilliQ water by dynamic light scattering (Malvern Zetasizer), and have been synthesized as described in \cite{Schneidegger2017}. The size response of the microgels was measured by dynamic light scattering (Malvern Zetasizer) from 23 to 40\textdegree{}C.
	The clusters were fabricated using sCAPA adapting a procedure described in previous work \cite{Ni2017}. Briefly, a 45 \textmu{}L droplet of a 0.1$\%$ v/v PS colloidal solution with 0.05 mM SDS (Sigma-Aldrich) and 0.005 wt\% Triton X-30 (Sigma-Aldrich) was confined between a template and a flat PDMS piece and dragged  at a speed of 3 \textmu{}m/s and a temperature of 25\textdegree{}C. The template contains an array of micro-sized rectangular traps of 2.2$\times$1.1 \textmu{}m\ and 0.5 \textmu{}m-depth over an area of 2 cm$^2$. After the droplet passes, traps are filled with one or two PS particles, in the central or lateral region of the template, respectively. The process is then repeated with an aqueous dispersion of PNIPAM-co-AAC microgels in 0.001$\%$ of Triton X-30, which are deposited in close contact with the PS particles. The local orientation of the droplet meniscus relative to the traps defines the number and position of deposited microgels as described in the main manuscript. The PS-microgel assemblies are sintered in the oven for 5 minutes at 70\textdegree{}C. Microgels deposited in the traps but not in contact with PS particles remain disconnected. Finally, the clusters are harvested by freezing a droplet of 5 \textmu{}L MilliQ water on the traps, and lifting it from the template, diluted in 3 \textmu{}L of MilliQ water; 7.4 \textmu{}L of the thawed particle solution are transferred to the sample cell for the experiments, without further treatment.

	\subsection{Experimental setup}
	The active colloidal clusters are imaged in a customized sample cell comprising two transparent electrodes, separated by an adhesive spacer with a 9 mm-circular opening and 0.12 mm height (Grace Bio-Labs SecureSeal, USA). The transparent electrodes were fabricated using 22 mm $\times$ 22 mm glasses (85--115 \textmu{}m-thick, Menzel Gl\"aser, Germany) coated via e-beam metal evaporation with 3 nm Cr and 10 nm Au (Evatec BAK501 LL, Switzerland), and plasma enhanced chemical vapor deposition with 10 nm of SiO$_2$ (STS Multiplex CVD, UK) to minimize particles sticking at the surface of the conductive glass slide. 
	The electrodes are connected to a function generator (National Instruments Agilent 3352X, USA) that applies the AC electric field, with a fixed frequency of 1 kHz and $V_\textrm{pp}$ between 4-5 V (32-42 V/mm). 
	Imaging was carried out in an inverted microscope (Axio Observer Z1) using 20$\times$ and 40$\times$ objectives (Zeiss). Image sequences (540 $\times$ 650, 16 bits) were taken with a sCMOS camera (Andor Zyla) at a frame rate of 10 fps and exposure time of 10 ms. The epifluorescence illumination source (North 89) was used to excite the PS particles at a $\lambda=580$ nm for imaging and local heating. Bright-field, transmission illumination was used in combination with epifluorescence to identify both components of the active clusters. Movies at high magnification (63x) were recorded with an inverted Eclipse Ti2-E (Nikon) microscope at 10 fps with 10 ms exposure time.
	
	\subsection{Image analysis}
	
	The collected image sequences were analyzed using custom particle tracking routines (Matlab) to follow the particle displacements and orientation as a function of time. Both instantaneous velocities and fitting of mean square displacements are used to extract quantitative data. In particular, a calibration of the local heating as a function of $\rho_{\rm FL}$ was obtained by measuring the Brownian translational diffusivity of the PS particles under different illumination intensities (Supplementary Fig.\ S3). A control experiment where particle the translational diffusivity was measured as a function of the global temperature imposed using a Peltier element at a fixed illumination is presented in Supplementary Figs.\ S3.
	As cross check, the translational diffusivities of non-fluorescent 1.93\ \textmu{}m diameter SiO$_2$ particles (Microparticles GmbH) were also measured using the same illumination conditions.
	
	\subsection{Dielectric Spectroscopy measurements}
	
	The dielectric properties of the PS particles and of the PNIPAM-co-AAC microgels were measured with a Novocontrol high-resolution dielectric analyzer (Alpha-A). The measurements were performed on a cell with a 6.45 mm gap distance between the electrodes enclosed by a Teflon cylinder, which was filled with the microgel suspension of 1 wt\%. The real and imaginary permittivity ($\epsilon'$,  $\epsilon''$), and conductivity ($\sigma'$, $\sigma''$) were determined over a wide frequency range ($10^2$-$10^7$ Hz) using temperatures below (25\textdegree{}C) and above (38\textdegree{}C) the VPTT of the microgels (Supplementary Fig.~S7). Moreover, the frequency was fixed at 1 kHz and the dielectric response of the microgel was evaluated increasing the temperature from 25\textdegree{}C to 40\textdegree{}C with a rate of  2\textdegree{}C/min (Supplementary Fig.~S2b).

	\bibliographystyle{naturemag}
	\bibliography{responsive}
	
\end{document}


\title{Supporting Information for: Reconfigurable Artificial Microswimmers with Internal Feedback}

\author{L. Alvarez}
\affiliation{Laboratory for Soft Materials and Interfaces, Department of Materials, ETH Zurich, 8093 Zurich, Switzerland}%
\author{M. A. Fernandez-Rodriguez}%
\affiliation{Biocolloid and Fluid Physics Group, Applied Physics Department, Faculty of Sciences, University of Granada, 18071 Granada, Spain}%
\author{A. Alegria}%
\author{S. Arrese-Igor}%
\affiliation{Centro de F\'isica de Materiales (SCIC-UPV/EHU), Materials Physics Center, 20018 San S\'ebastian, Spain}
\author{K. Zhao}
\affiliation{Laboratory for Soft Materials and Interfaces, Department of Materials, ETH Zurich, 8093 Zurich, Switzerland}%
\author{M. Kr\"oger}
\affiliation{Polymer Physics, Department of Materials, ETH Zurich, 8093 Zurich, Switzerland}%
\author{Lucio Isa}
\affiliation{Laboratory for Soft Materials and Interfaces, Department of Materials, ETH Zurich, 8093 Zurich, Switzerland}%
\email{lucio.isa@mat.ethz.ch}
	\maketitle
	
\onecolumngrid
\parindent 0mm
\bigskip

\textbf{Movie S1}. Reconfiguration upon local temperature change of the PS-microgel assembly via fluorescence illumination cycles (on and off).
\medskip

\textbf{Movie(s) S2-3}. Dynamical control of active PS-microgel assemblies under AC electric field ($\omega=1$ kHz and  $V_{\rm pp}=4$ V) during various    ON and OFF cycles of fluorescent illumination. Only PS fluorescent particles are visible (40x magnification in different ROIs).
\medskip

\textbf{Movie S4}. Reversal of swimming direction of and active PS-microgel dumbbell due to a increase of local temperature when illuminating at high fluorescence intensity (60x magnification).
\medskip

\textbf{Movie S5}. L-shape active motion at different illumination conditions exhibiting chirality inversion upon self-reconfiguration.
\bigskip

\parindent 4mm 
	
	\begin{figure}[H]
		\centering
		\includegraphics[width=0.9\columnwidth]{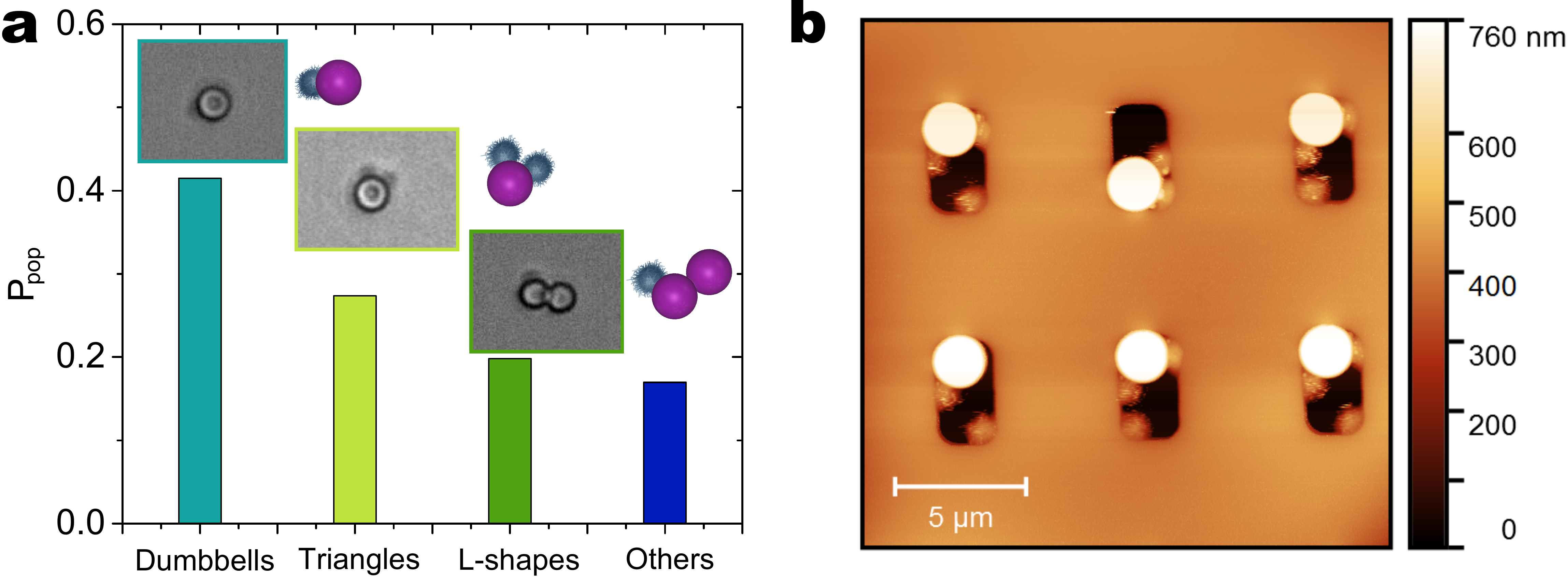}
		\caption{\label{figS1}\textbf{a}, Histogram of particle population after sCAPA deposition and transferring to an experimental cell. Insets represent optical microscopy images of the dominant particle populations: dumbbells, triangles and L-shapes with their corresponding schemes. The histogram is done by classifying 190 particles after transferring. \textbf{b}, AFM image (in water) of the triangular PS-microgel clusters inside the sCAPA template before harvesting, showing some of the assembly defects.}
		\label{fig:figS1}
	\end{figure}
	
	\clearpage
	
	\begin{figure}[H]
		\centering
		\includegraphics[width=\columnwidth]{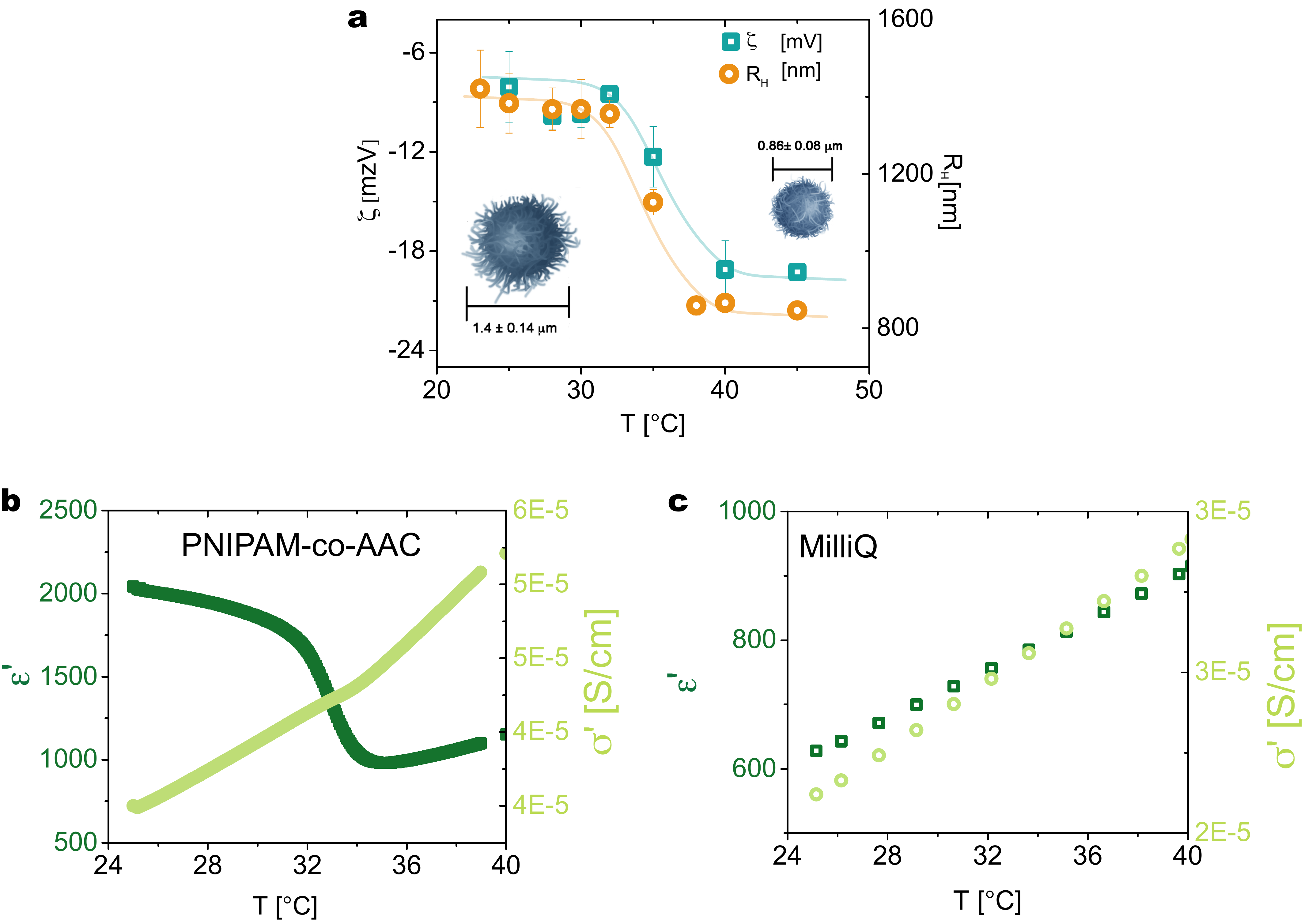}
		\caption{\label{figS2}Temperature dependence of microgel properties. \textbf{a}, Zeta potential ($\zeta_\textrm{\textmu{}gel}$, green) and hydrodynamic diameter ($D_\textrm{\textmu{}gel}$, orange) vs. $T$. The insets schematically represent the swollen (left) and collapsed (right) microgel before and after its VPTT, respectively. The lines are a guide to the eye. \textbf{b}, Dielectric constant ($\epsilon'$, dark green) and conductivity ($\sigma'$, light green) vs. $T$ of a PNIPAM-co-AAC microgel suspension (1 wt \%) after subtracting the electrode polarization and MilliQ water contributions at $f$ = 1 kHz. \textbf{c}, $\epsilon'$ and $\sigma'$ for  MilliQ Water as a function of $T$ at the same frequency $f$}
		\label{fig:figS2}
	\end{figure}

	\begin{figure}[H]
		\centering
		\includegraphics[width=0.95\columnwidth]{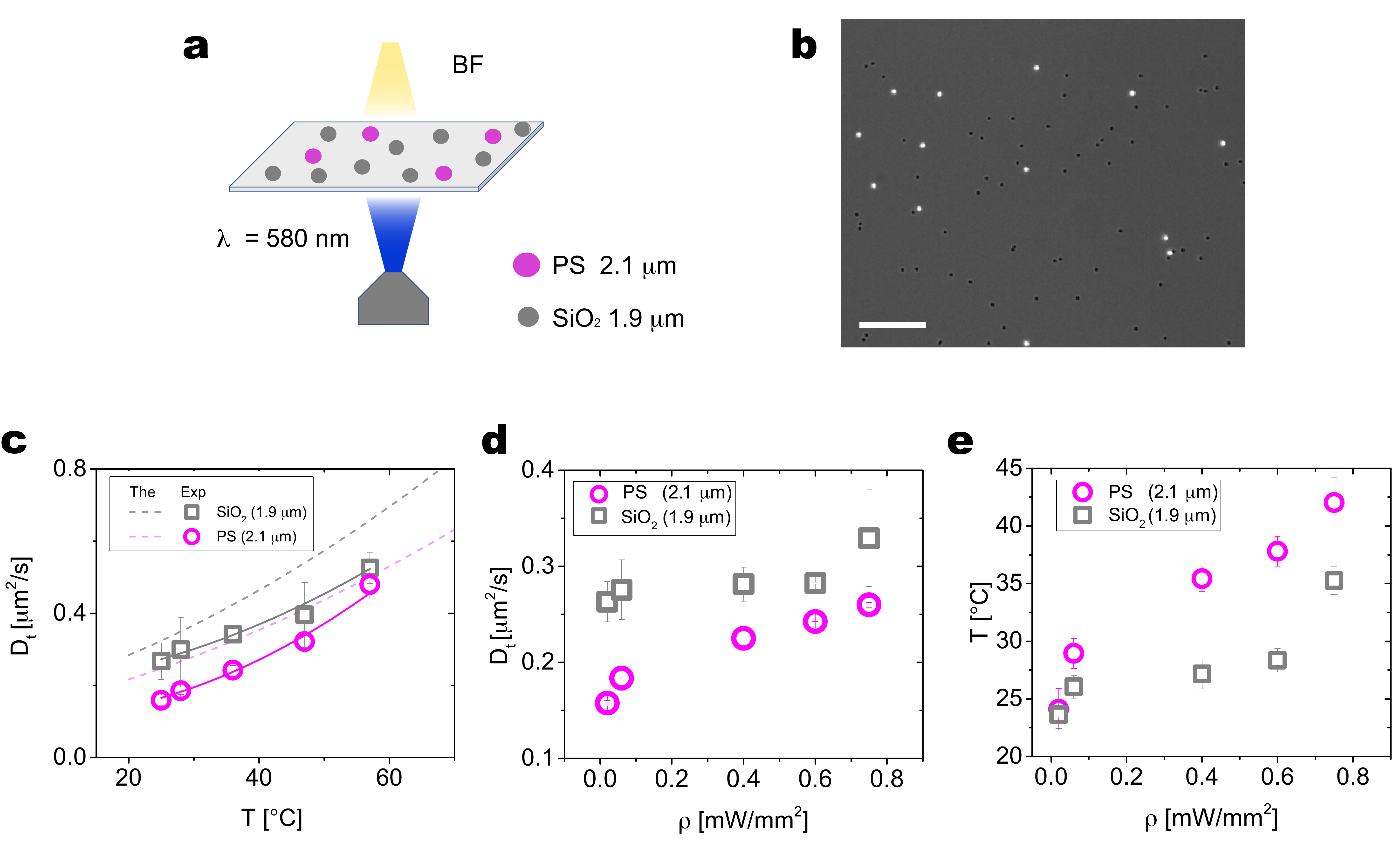}
		\caption{\label{figS3}Calibration of PS particle heating vs. fluorescence illumination power density. \textbf{a-b}, Scheme \textbf{a} and optical micrograph \textbf{b} of the experiment. By mixing epifluorescence ($\lambda = 580$ nm) and transmission illumination, PS (bright) and SiO$_2$ (dark) particles in water can be simultaneously tracked under the same conditions.  In all cases the error bars represent the standard deviation of 120 measured particles. Scale bar: 20 \textmu{}m. \textbf{c}, Translational diffusion coefficient $D_t$ for PS (pink) and SiO$_2$ particles (gray) as a function of $T$, globally imposed by a Peltier element. The dashed line show the theoretical value in bulk for each particle. The solid lines are 2nd order polynomial fittings of the data as $D_t = A + BT + CT^2$.  The error bars represent the standard deviation of 120 measured particles.
			\textbf{d}, $D_t$ measured values imposing the $\rho_{\rm FL}$ intensities used in the experiments. 
			\textbf{e}, Conversion between $\rho_{\rm FL}$ and $T$ via the measured values of $D_t$ . The extrapolation of $T$ is done from the fitted expression using the diffusivities of both particles under fluorescent illumination from \textbf{d}. The data in \textbf{e} shows that over the range $\rho_{\rm FL}$ $<0.6$ mW/mm$^2$, no strong change in diffusivity is seen for the SiO$_2$ particles. At illumination power densities $\rho_{\rm FL} >0.6$ mW/mm$^2$ the fluorescent light also induces strong global heating of the sample cell. The error propagation throughout the extrapolation has been included in the error bars of $T$.}
	\end{figure}
	

	\begin{figure}[H]
		\centering
		\includegraphics[width=\columnwidth]{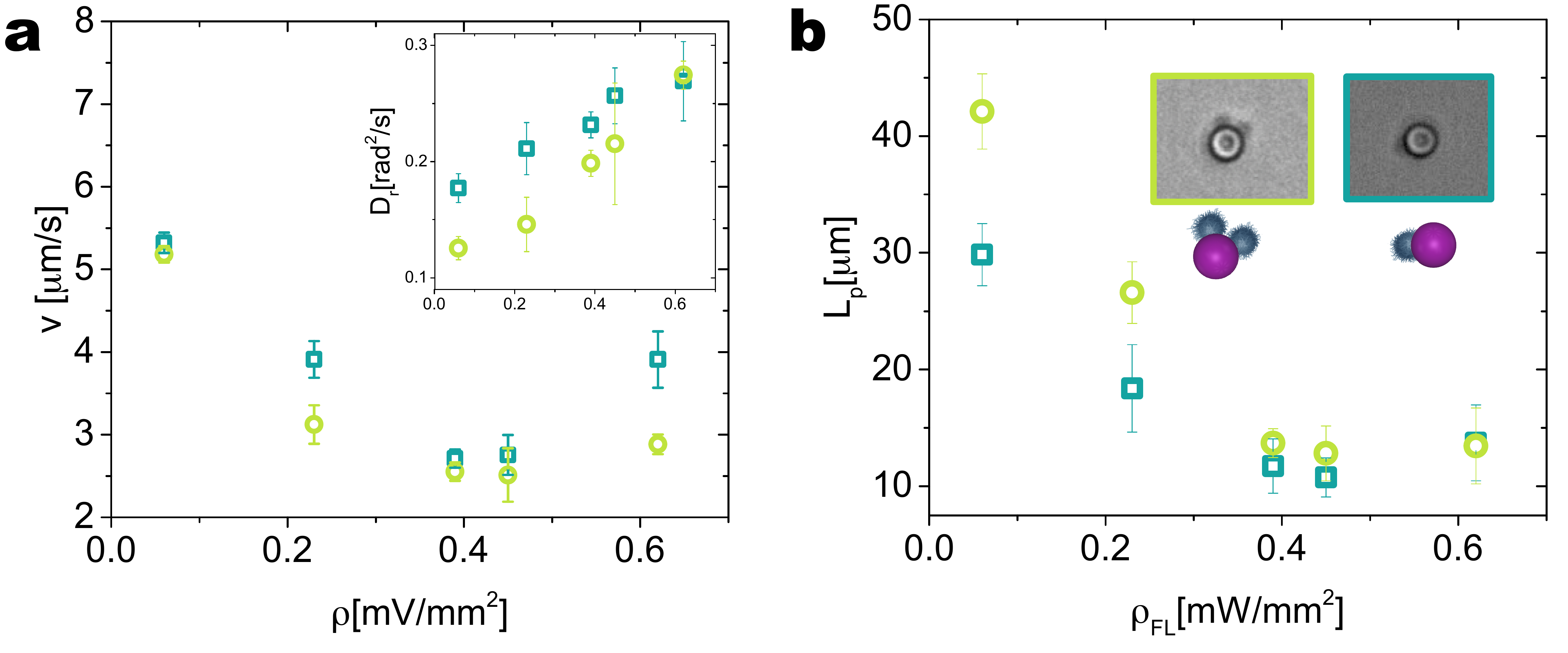}
		\caption{\label{figS4}\textbf{(a)} Swimming velocities $v$ and and rotational diffusivity $D_r$ of dumbbells (squares) and triangles (circles) as a function of $\rho_{\rm FL}$. \textbf{(b)} Persistence length of the trajectories $L_p = v / D_r$ for each cluster shape (same symbols). The inset depicted a optical microscopy picture and its corresponding schematic for each active cluster.  The dynamics were obtained over 43 particles for the dumbbells and over 14 particles for the triangles.}
	\end{figure}

	\begin{figure}[H]
		\centering
		\includegraphics[width=00.7\columnwidth]{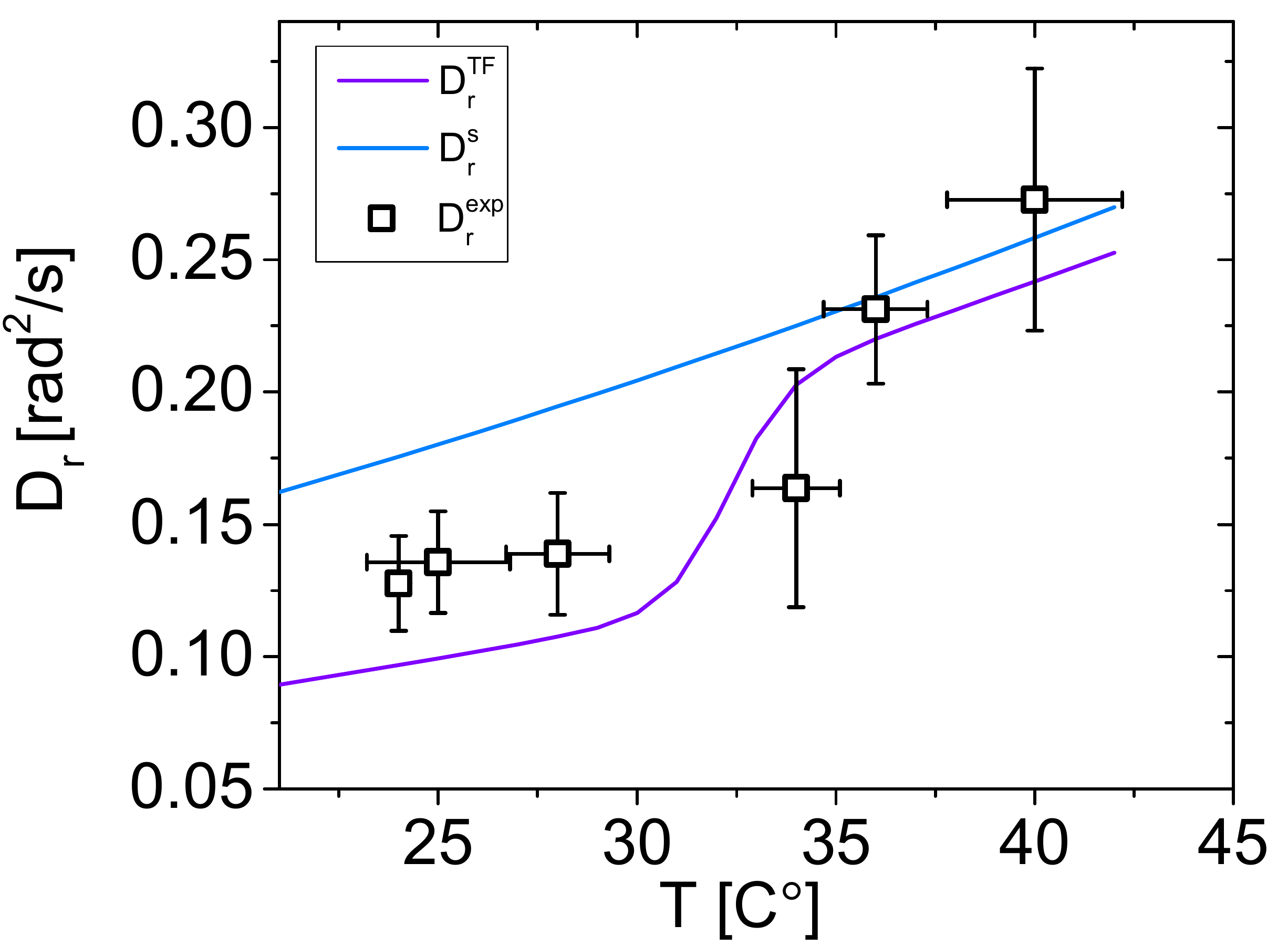}
		\caption{\label{figS5}Dumbbell's rotational diffusion. \textbf{(a)} Comparison between experimental data ($D_r^\textrm{exp}$, black squares) and theory for a torque free dumbbell ($D_r^\textrm{TF}$, purple line), and for a sphere ($D_r^s$, blue line). Error bars correspond to the standard deviation of $D_r^\textrm{exp}$ (y-error bar) and to the uncertainty in measuring $T$ (x-error bar).}
	\end{figure}
	
	\begin{figure}[H]
		\centering
		\includegraphics[width=0.8\columnwidth]{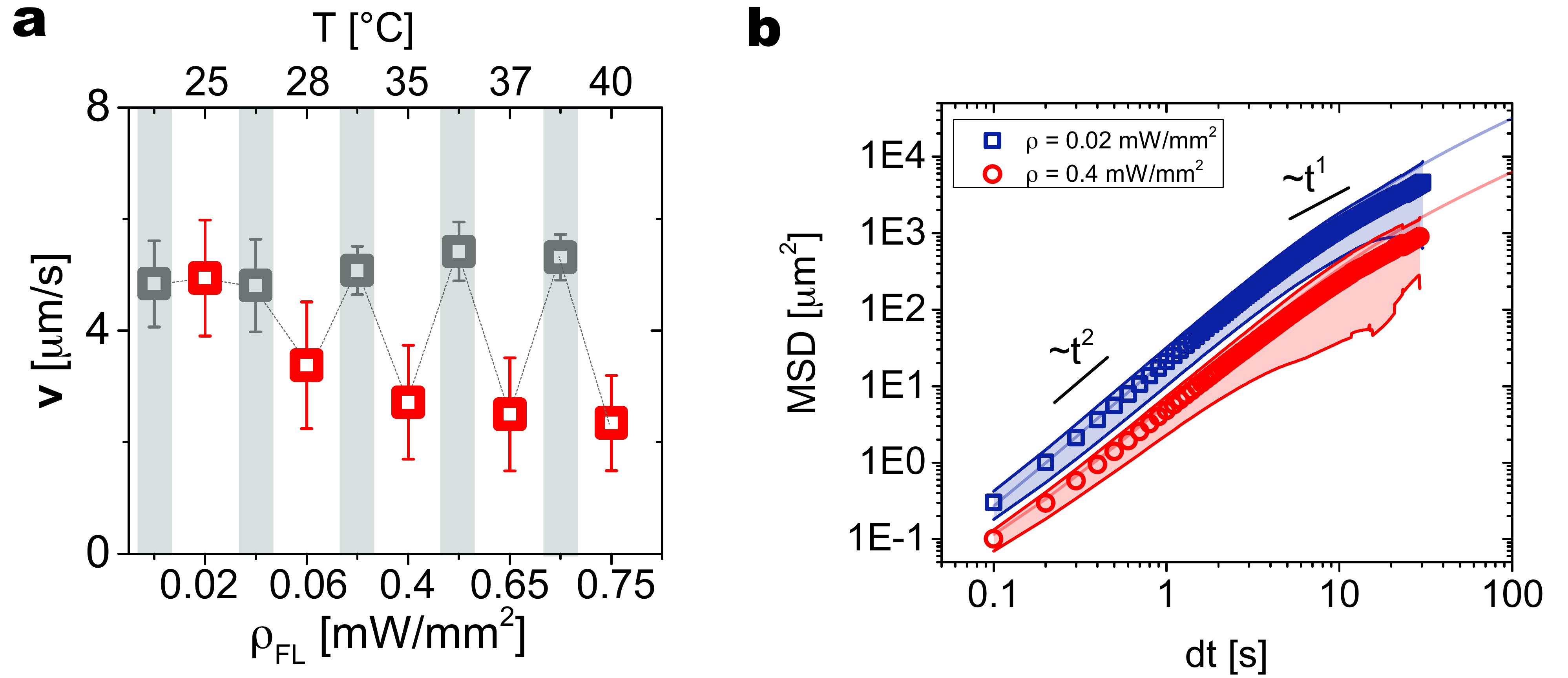}
		\caption{Dumbbell's velocity response. \textbf{(a)} Dumbbell swimming velocity (absolute value) during illumination cycles of no fluorescent light (OFF - gray squares) and different  values of $\rho_{\rm FL}$ (red squares) corresponding to different local $T$ (top x-axis). Each data point was collected after a 5-second equilibration. \textbf{(b)} Mean square displacements of active dumbbells at two values of $\rho_{\rm FL} =0.02$ and 0.4 mW/mm$^2$, respectively below and above the microgel VPTT, showing that the distinct transition from ballistic $\sim t^2$ to diffusive $\sim t^1$ can be tuned by illumination. Error bars in all cases indicate the standard deviation of values calculated over 150 particles.}
		\label{fig:Labview}
	\end{figure}

	\section{Measurement of the dielectric properties of PNIPAM-co-Aac microgels}

\begin{figure}[H]
		\centering
		\includegraphics[width=\columnwidth]{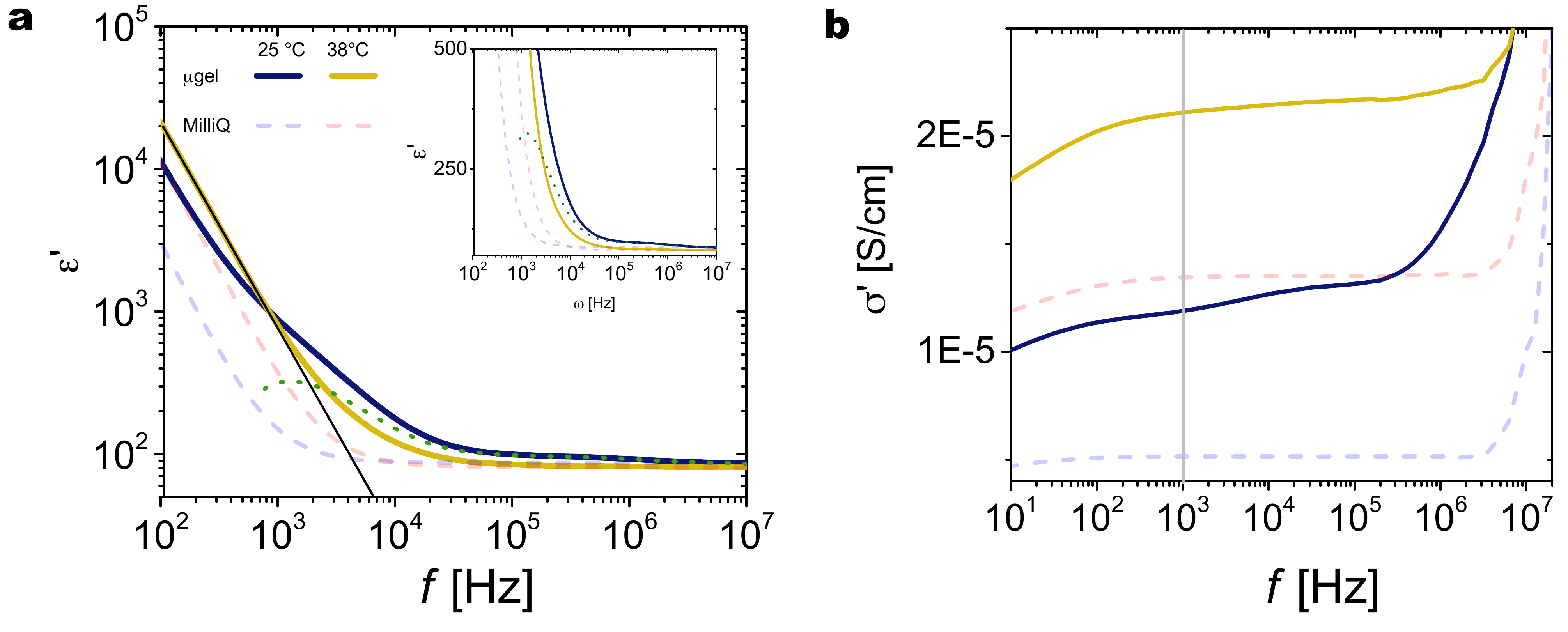}
		\caption{Dielectric spectroscopy measurements.  Dependence of \textbf{(a)} $\epsilon'$ and \textbf{(b)} $\sigma'$  with frequency $f$ for an aqueous suspension of microgels (solid lines) and water (dashed lines) at $25$\textdegree{}C (blue) and $38$\textdegree{}C (yellow). The corrected $\epsilon'$ data of the microgel at $25$\textdegree{}C from the electrode polarization effect is represented as the dashed green line. The black solid line represent the fitting for the electron polarization effect.}
		\label{fig:dielectric}
	\end{figure}

	The dielectric properties ($\epsilon'$ and $\sigma'$) of the microgels within a broad range of frequencies were quantified using dielectric spectroscopy (Fig.\ \ref{fig:dielectric}).
	We performed two sets of measurements, one with a microgel suspension of 0.1 wt\% and another one simply using MilliQ water as control measurement. Qualitatively, we observe a difference from the solution below ($25$\textdegree{}C) and above ($38$\textdegree{}C) the VTPP of the microgel. Between frequencies $f= 1-10$ kHz at $25$\textdegree{}C (blue line) there is a first relaxation associated to the charge mobility within the core of the microgel, which disappears when increasing temperature $38$\textdegree{}C (yellow line), indicating the drastic variation of the dielectric constant of the microgel above the VTTP.
	The relative values of $\epsilon'$ of the microgels ($\epsilon_p$) were obtained considering that $\epsilon_\textrm{exp} = \epsilon_p + \epsilon_m + \epsilon_\textrm{ep} $ \cite{Mohanty2016}, where $\epsilon_\textrm{exp}$ is the raw experimental values obtained from the measurements, $\epsilon_\textrm{ep}$ is the contribution from the electrode polarization, and $\epsilon_{m}$ is dielectric constant of medium, in this case 79.1 (for water at $f$= 1 kHz and $25$\textdegree{}C) \cite{Archer1990}). In particular, within the window of our interest frequencies ($f$= 1kHz), the effect of electrode polarization can be fit by the initial power-law decay of $\epsilon'$ using the expression $m\omega^{-n}$ \cite{Su2014}. First, we fitted the raw permittivity data and determined the values of $m$ and $n$. The EP effect was then subtracted from the raw permittivity data at each frequency by use of $m$ and $n$. The difference between the experimental data and the corrected data at $1$ kHz is used to calculate the corrected values for the data as a function of temperature shown in Fig.\ \ref{fig:figS2}. The real part of the conductivity $\sigma'$ is then recalculated from the new corrected values of the $\epsilon'$ \cite{Deshmukh2017}.

	\section{Calculation of the T-dependent propulsion speed for a PS-microgel dumbbell }

	\begin{figure}[H]
		\centering
		\includegraphics[width=0.8\columnwidth]{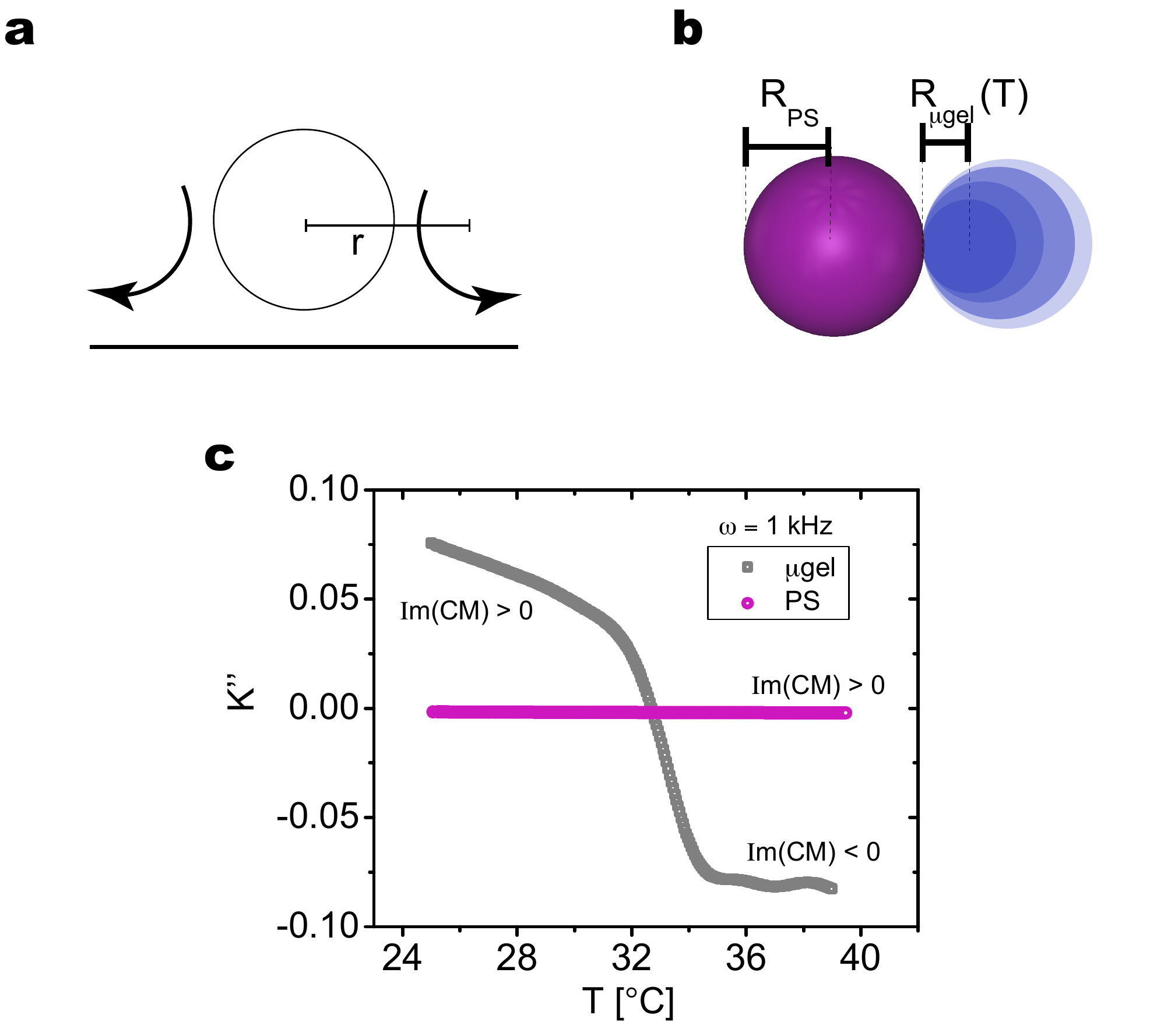}
		\caption{\textbf{(a)} Schematic representation of the EHD flows around a single sphere (arrows). $r$ represents the distance at which the magnitude of the EHD flow is evaluated. \textbf{(b)} Scheme of a PS-microgel dumbbell. The EHD flows of each lobe are evaluated at a distance corresponding to $r(T)=R_\textrm{PS} + R_\textrm{\textmu{}gel}(T)$. \textbf{(c)} Temperature dependence of $K''$ calculated from Eq.\ \eqref{eq:MSDlong} for the microgel (gray) and a PS sphere (pink) at a fixed frequency of 1 kHz. }
		\label{fig:rings}
	\end{figure}
	
	The theoretical prediction of the EHD flows generated by a single sphere under an applied AC voltage $V_{\rm pp} e ^{-j\omega t}$ has been studied and derived by  S. Wu \emph{et al.} in previous works \cite{Ma2015, Ma2015a}. The EHD flows depend on $D$ (the diffusion coefficient of the ions), $\kappa^{-1}$ (the Debye length) and $H$ (the half-separation between the electrodes, 60 \textmu{}m in our case). Therefore, the velocity of the EHD flow $U_i$ around a single particle of radius $R_i$ at a given reference distance $r_i$ at which the EHD is evaluated, within 2 electrodes of distance $h=2H$ can be calculated using
	\begin{equation}
		\label{eq:EHD} U_i = \frac{C K''}{\mu}\frac{3(r_i/R_i)}{2 \left[ 1+ (r_i/R_i)^2\right]^{5/2}},
		\qquad
		C = \epsilon \epsilon_0 \left( \frac{V_p}{2H} \right)^2 \frac{\alpha^2}{1+\alpha^2} \frac{\kappa D}{\omega}
	\end{equation}
	%
	%
	where 
	$\alpha = \omega H/ \kappa D$ and $K''$ is the imaginary part of the Clausius-Mossotti factor \cite{CMFactor2017}; $\mu$ is the fluid viscosity. 
	
	To evaluate the EHD for each single lobe, we consider $r_i$ as the distance from the center of the evaluated lobe to the center of the adjacent one (Fig. \ref{fig:rings}). Therefore, as the radius of the microgel $R_\textrm{\textmu{}gel}$ varies with temperature, $r_i$ also changes. Moreover, the dielectric properties of the PS sphere and the microgel have a different T-dependent behavior. In particular, they enter the formula above through $K''$ as
	\begin{equation}
		\label{eq:MSDlong} K''= \frac{\omega (\sigma_m - \sigma_p)(\epsilon_p+2\epsilon_m) - (\epsilon_p - \epsilon_m)(\sigma_p + 2\sigma_m)}{\omega^2(\epsilon_p+2\epsilon_m)^2 + (\sigma_p + 2\sigma_m)^2},
	\end{equation} 
	where $\sigma_m$ and $\epsilon_m$ are the dielectric properties of the media, and  $\sigma_p$ and $\epsilon_p$ the dielectric properties of the particles. The behavior of the $K''$ factor depends on the values of $\sigma'$ and $\epsilon'$ of each particle. For the calculation of the $K''$ of the microgel we use the corrected experimental values as explained in previous section and we use tabulated values for the PS particles \cite{ERMOLINA2005419}.

	\section{Experimental obervation of EHD flows around the dumbbell}
	
\begin{figure}[H]
		\centering
		\includegraphics[width=0.75\columnwidth]{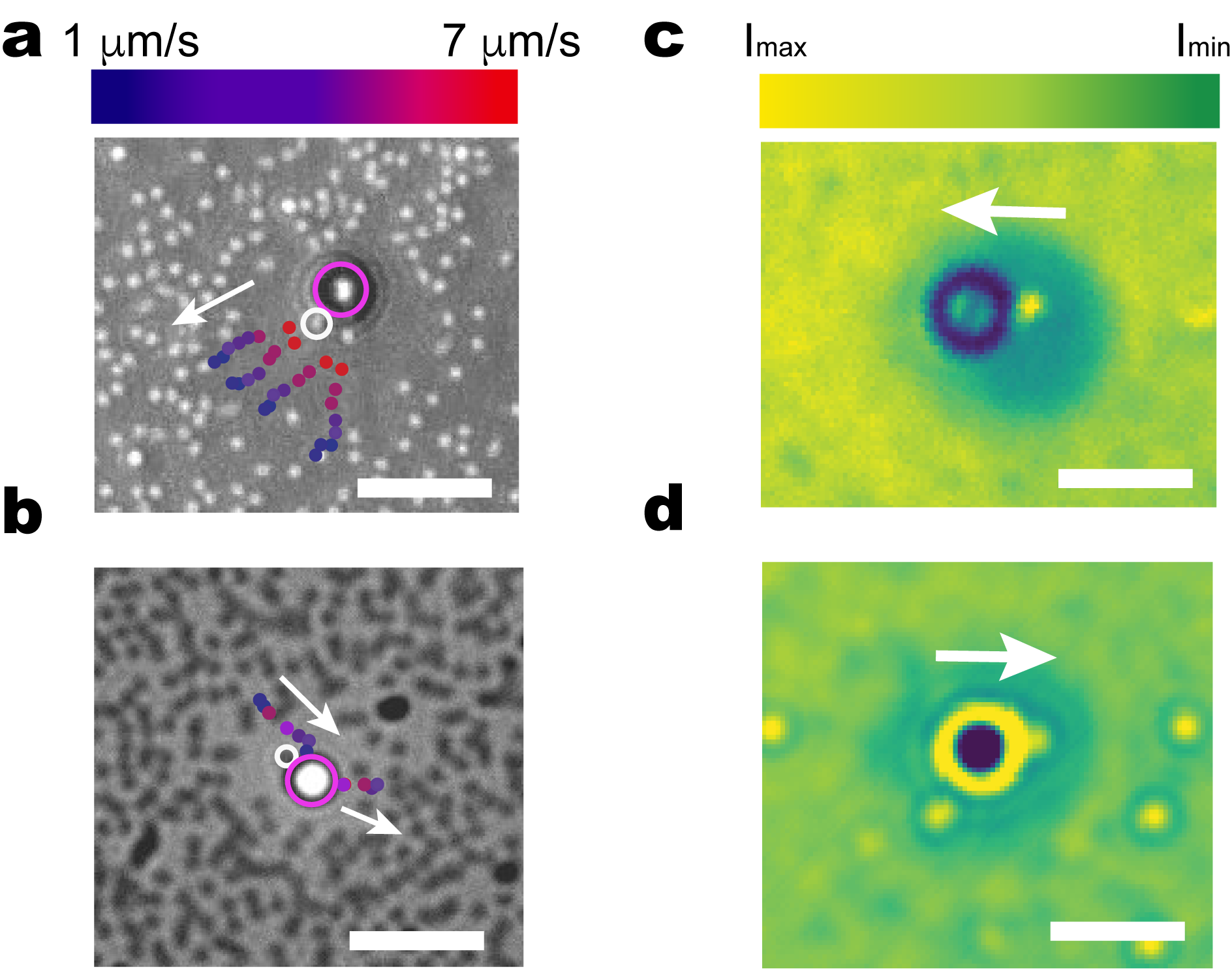}
		\caption{ Experimental characterization of EHD flows \textbf{(a-b)} Trajectories of PS 700 nm tracer particles around the dumbbells under an external AC electric field of 5 V and 1 kHz at low (0.02 mW/mm$^2$) \textbf{(a)} and high (0.6 mW/mm$^2$) \textbf{(b)} illumination conditions $\rho_\textrm{FL}$. The color coding represent the instantaneous velocity of the tracers. The arrows represent the direction in which the tracers are moving. Scale bars represent 6 \textmu{}m \textbf{(c-d)} Tracers intensities sum over 60 s at  low (0.02 mW/mm$^2$) \textbf{(c)} and high (0.6 mW/mm$^2$) \textbf{(d)} $\rho_\textrm{FL}$. The color coding depicts the intensity range from $I_\textrm{max}$ (bright green) to $I_\textrm{min}$ (dark green). The white arrows indicate the swimming direction for each condition Scale bars represent 4 \textmu{}m.}
		\label{fig:EHDscheme}
	\end{figure}

	The direction of the EHD flows can be experimentally observed by using small fluorescent tracers. The tracers at each side of the dumbbell might be ejected (repulsive) or taken in (attractive), revealing the direction of the EHD (Fig. \ref{fig:EHDscheme}). At low light intensities , we observe that tracers are strongly ejected away from the microgel, while the same phenomenon occurs with  a much lower intensity in the proximity of the PS lobe. This indicates that, albeit both flows are repulsive, they are highly asymmetric and the overall behavior is dominated by the flow around the microgel surface, causing the dumbbell to propel with the PS lobe in front. When the microgel shrinks at higher light intensities, the sign of its EHD flow changes, and the flow slightly reduces. At the same time, the sign of the PS EHD remains the same but increases slightly due to local viscosity decrease. The combination of the latter effects causes an inversion of the propulsion direction of the dumbbell,  swimming with the microgel in front.

	\bibliographystyle{naturemag}
\bibliography{responsive}